\pdfoutput=1
\documentclass[doublecol]{epl2}
\usepackage{amsmath,amssymb}
\usepackage[numbers,sort&compress]{natbib}
\usepackage{url}
\usepackage{hyperref}

\makeatletter
\def\NAT@spacechar{~}
\makeatother
\def\vec#1{\mathbf{#1}}
\let\hati\hat
\def\hat#1{\hati{\mathbf{#1}}}
\def\dd#1{\mathop{}\!\mathrm{d}{#1}}

\def\TPI{Institut f\"ur Theoretische Physik I --- Ruhr-Universit\"at Bochum, 44801 Bochum, Germany}
\def\TPIV{Institut f\"ur Theoretische Physik IV --- Ruhr-Universit\"at Bochum, 44801 Bochum, Germany}
\author{%
    Jeremiah Lübke\,%
    \inst{1}\thanks{Email: \email{jeremiah.luebke@rub.de}}
    \and
    Frederic Effenberger\,%
    \inst{1,2}
    \and
    Mike Wilbert\,%
    \inst{1}
    \and
    Horst Fichtner\,%
    \inst{2}
    \and
    Rainer Grauer\,%
    \inst{1}
}
\shortauthor{J. Lübke \etal}
\institute{%
    \inst{1} \TPI \\
    \inst{2} \TPIV
}
\title{Towards Synthetic Magnetic Turbulence with Coherent Structures}
\date{\today}

\begin{document}
\abstract{%
Synthetic turbulence is a relevant tool to study complex astrophysical and space plasma environments inaccessible by direct simulation. 
However, conventional models lack intermittent coherent structures, which are essential in realistic turbulence.
We present a novel method featuring coherent structures, conditional structure function scaling and fieldline curvature statistics comparable to magnetohydrodynamic turbulence.
Enhanced transport of charged particles is investigated as well.
This method presents significant progress towards physically faithful synthetic turbulence.}

\maketitle

\section{Introduction}
Turbulence plays a key role in astrophysical and space plasma environments~\cite{cho_GenerationMagneticFields_2000,ryu_MagneticFieldsLargeScale_2012,hosking_CosmicvoidObservationsReconciled_2023,manzini_SubionScaleTurbulenceDriven_2023,bruno_SolarWindTurbulence_2013,engelbrecht_TheoryCosmicRay_2022}.
However, due to the high computational cost of direct approaches, the effect of turbulence in such environments is difficult to study.
This obstacle is often mitigated by splitting the magnetic field in a large-scale coherent component with an analytic description and a small-scale turbulent component, modelled as a Gaussian random field~\citep[e.g.,][]{tautz_SimulatingHeliosphericSolar_2011,beck_NewConstraintsModelling_2016}.
Such Gaussian random fields can be easily synthesized as a superposition of plane waves with random phases and a prescribed energy spectrum.
The transport of energetic charged particles through such fields has been extensively studied~\citep[e.g.,][]{neuer_DiffusionTestParticles_2006,subedi_ChargedParticleDiffusion_2017,dundovic_NovelAspectsCosmic_2020,mertsch_TestParticleSimulations_2020,reichherzer_EfficientMicromirrorConfinement_2023}.

However, Gaussian random fields can only provide a low-order approximation of magnetic turbulence, neglecting any structure beyond two-point correlations captured by the energy spectrum.
They do not exhibit intermittency as observed in first-principles turbulence~\citep[e.g.,][]{she_UniversalScalingLaws_1994,grauer_ScalingHighorderStructure_1994,roberts_ScaleDependentKurtosisMagnetic_2022,gomes_OriginMultifractalitySolar_2023}.
Intermittency was studied in the context of hydrodynamic synthetic turbulence models already by Juneja \etal~\cite{juneja_SyntheticTurbulence_1994}, and its impact on charged particle transport more recently by Pucci \etal~\cite{pucci_EnergeticParticleTransport_2016} and Shukurov \etal~\cite{shukurov_CosmicRaysIntermittent_2017}, finding faster diffusion in structured magnetic fields.

Up until today, there have been several models for synthetic hydrodynamical and magnetohydrodynamical~(MHD) turbulence published,
such as the $p$-model on a discrete three-dimensional wavelet space by Malara \etal~\cite{malara_FastAlgorithmThreedimensional_2016},
the minimal multiscale Lagrangian map by Rosales and Meneveau~\cite{rosales_MinimalMultiscaleLagrangian_2006},
which was applied to MHD turbulence by Subedi \etal~\cite{subedi_GeneratingSyntheticMagnetic_2014},
a stochastic integral based on the lognormal-model including asymmetric velocity increment statistics by Pereira \etal~\cite{pereira_DissipativeRandomVelocity_2016},
and its application to MHD turbulence by Durrive \etal~\cite{durrive_MagneticFieldsMultiplicative_2020}.
All of these models produce three-dimensional, divergence-free vector fields with intermittent statistics, but without coherent geometric features.
Recently, Durrive \etal~\cite{durrive_SwiftGeneratorThreedimensional_2022} presented a model which embeds Archimedean spirals into a random lognormal vector field.
The continuous wavelet cascade by Muzy~\cite{muzy_ContinuousCascadesWavelet_2019} addresses broken stationarity of discrete wavelet cascades.
Related works were recently published by Li \etal~\cite{li_SyntheticLagrangianTurbulence_2023} and Robitaille \etal~\cite{robitaille_StatisticalModelFilamentary_2020}.

Standard tools of validating synthetic models are the energy spectrum and statistics of field increments.
Taken without further decomposition, these quantities provide a global picture of the vector field, hiding the intricate local geometry of magnetic turbulence~\citep[see also][]{schekochihin_MHDTurbulenceBiased_2022,groselj_KineticTurbulenceAstrophysical_2019}.
A useful quantity in this regard is the fieldline curvature, which has recently been shown by Kempski \etal~\cite{kempski_CosmicRayTransport_2023} and Lemoine~\cite{lemoine_ParticleTransportLocalized_2023} to play a key role in the transport of charged particles in magnetic turbulence.
Fieldline curvature has previously been discussed in the context of turbulent dynamos~\citep{schekochihin_SimulationsSmallScale_2004}, as well as hydrodynamic~\citep{bentkamp_StatisticalGeometryMaterial_2022,qi_FoldingDynamicsIts_2023,xu_CurvatureLagrangianTrajectories_2007} and magnetohydrodynamic turbulence~\citep{yang_RoleMagneticField_2019,yuen_CurvatureMagneticField_2020}.

In this letter we present progress towards a model for synthetic magnetic turbulence featuring intermittent coherent structures.
We implement the model as a fast algorithm, which produces a random three-dimensional divergence-free vector field, resembling a turbulent magnetic field~$\vec{b}(\vec{x})$~\footnote{The code is publicly available at \url{https://github.com/jerluebke/synth-mag-turb} and archived at \href{https://doi.org/10.5281/zenodo.10515965}{doi:10.5281/zenodo.10515965}. A pseudocode listing is provided as supplementary material.}.
The model is a combination of a continuous cascade~\citep{muzy_ContinuousCascadesWavelet_2019} and the minimal multiscale Lagrangian map~\citep{rosales_MinimalMultiscaleLagrangian_2006,subedi_GeneratingSyntheticMagnetic_2014}.
Additionally, we propose a set of quantities to assess the physical fidelity of synthetic turbulence models, consisting of the energy spectrum, conditional structure function scaling, the fieldline curvature distribution and running diffusion coefficients of charged test particles.
Based on these quantities, we compare the proposed synthetic turbulence model with an incompressible resistive MHD turbulence simulation and an intermittent synthetic turbulence model without coherent structures.
We also consider the phase-randomized counterparts of the three turbulence models to account for differences in the energy spectra.
We conclude by explaining the shape of the MHD fieldline curvature distribution by means of a weighted sum of Gaussian components.

\begin{figure*}
    \centering
    \includegraphics{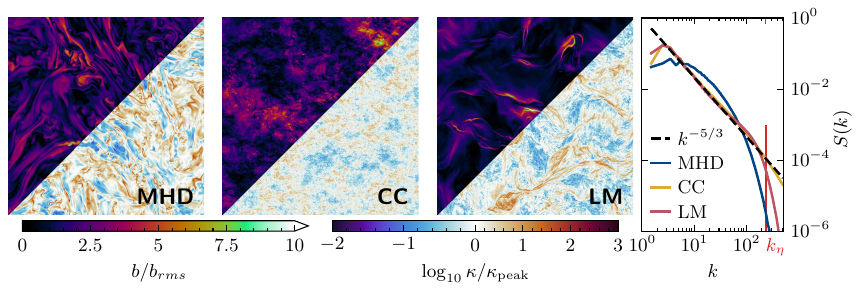}
    \caption{%
    Slice plots of magnetic field strength~$b/b_{rms}$ and logarithm of fieldline curvature~$\kappa/\kappa_\mathrm{peak}$ for three models of magnetic turbulence --- Magnetohydrodynamics (MHD), Continuous Cascade (CC) and Lagrangian Mapping (LM).
    Energy spectra of the models are plotted in the right-most panel and compared with a~$-5/3$rd scaling.
    The dissipation wavenumber of the MHD simulation~$k_\eta$ is indicated on the abscissa.
    The fieldline curvature is defined as~$\kappa=\|\hat{b}\cdot\nabla\hat{b}\|$, with~$\hat{b}=\vec{b}/\|\vec{b}\|$, and normalized by the most frequent value $\kappa_\mathrm{peak}$.
    }
    \label{fig:spectrum}
\end{figure*}

\section{Methods}
We start by extending the continuous cascade in wavelet space~\citep{muzy_ContinuousCascadesWavelet_2019} to three-dimensional divergence-free vector fields.
The continuous cascade at scale~$l$ and position~$\vec{x}$ is represented by a log-infinitely divisible process~$e^{\omega_l(\vec{x})}$, which gives the scale- and position-dependent intensity of a vector field~$\vec{v}(\vec{x})$.
This field is obtained by a vector-valued wavelet transform of~$l^He^{\omega_l}$ over the inertial range scales~$l_\mathrm{min}<l<l_0$ as
\begin{equation}
    \vec{v}(\vec{x})=\nabla\times A\int_{l_\mathrm{min}}^{l_0}l^{H-d-1}\big(e^{\omega_l}R_l\ast l{\psi}_l\hat{z}\big)(\vec{x})\dd{l}.
    \label{eq:cascade}
\end{equation}
The slope of the energy spectrum is given by~$-2H-1$,
torodial wavelets~$\nabla\times\big(l\psi_l(\vec{x})\hat{z}\big)$ with~$\psi(\vec{k})=-k^2e^{-k^2}$ and~$\psi_l(\vec{x})=\psi(\vec{x}/l)$ ensure the zero-divergence condition~$\nabla\cdot\vec{v}=0$,
a random rotation field~$R_l(\vec{x})$ with correlation length~$l$ ensures proper isotropization of these wavelets,
and the numerically determined constant~$A$ normalizes the field to~$\langle v^2\rangle=1$.
Further, the curl can be moved in front of the spatial convolution operator~$\big(f\ast \nabla\times g\big)(\vec{x})=\nabla\times\int_{\mathbb{R}^3}f(\vec{y})g(\vec{x}-\vec{y})\dd{\vec{y}}$ and the scale integral~$\int\cdots\dd{l}$,
thus allowing us to express the field in terms of a vector potential~$\vec{v}=\nabla\times\vec{a}$.

The infinitely divisible process~$\omega_l(\vec{x})$ is defined on cone-like subsets of the position-scale half-space~$\mathbb{R}^d\times\mathbb{R}_{>0}$ equipped with the measure~$l^{-d-1}\dd{\vec{x}}\dd{l}$ and has a cumulant generating function~$\phi(q)$.
Thus, the moments of the intensity process can be computed as~$\langle e^{q\omega_l}\rangle=(l_0/l)^{c_d\phi(q)}\ \forall q\in\mathbb{N}$, where~$c_d=2^{-d}\pi^{d/2}/\Gamma(d/2+1)$ comes from the scale-space cone volume.
We consider a Gaussian distribution with~$\phi(q)=\mu/2(p^2-p)$ with intermittency parameter~$\mu$, in which case~$e^{\omega_l}$ corresponds to the Gaussian multiplicative chaos employed in~\cite{pereira_DissipativeRandomVelocity_2016} and related works.

As shown below, the vector field given by Equation~(\ref{eq:cascade}) exhibits (isotropic) anomalous scaling, but lacks coherent features, so we introduce advective structures
by adapting the minimal multiscale Lagrangian map~(MMLM) for MHD~\citep{subedi_GeneratingSyntheticMagnetic_2014} to our framework.
In short, the MMLM procedure considers only the advective part of the magnetic field evolution equation, i.e.~$\left(\partial_t+\vec{u}\cdot\nabla\right)\vec{b}=0$, which can be solved at time~$\tau$ with the Lagrangian ansatz~$\vec{b}(\vec{x}_\tau,\tau)=\vec{b}(\vec{x}_0,0)$ and the linearized solution~$\vec{x}_\tau=\vec{x}_0+\tau\vec{u}(\vec{x}_0,0)$.
Usually, two initially Gaussian random vector fields representing~$\vec{u}(\vec{x})$ and~$\vec{b}(\vec{x})$ are deformed on successively finer scales~$l_i$ by applying the linearized Lagrangian solution with~$\tau\propto l_i$ to the underlying regular grid, as described in the references.

In our framework, we generate two independent random fields according to Equation~(\ref{eq:cascade}), which are represented as vector potentials and represent the velocity field~$\vec{u}(\vec{x})=\nabla\times\vec{a}^{\vec{u}}(\vec{x})$ and the magnetic field~$\vec{b}(\vec{x})=\nabla\times\vec{a}^{\vec{b}}(\vec{x})$.
When generating~$\vec{u}(\vec{x})$ scale-by-scale, we make use of the discretization of the scale integral in Equation~(\ref{eq:cascade}) as a sum~$\int_{l_\mathrm{min}}^{l_0}\cdots\dd{l}\approx\sum_{l_i=l_0}^{l_\mathrm{min}}\cdots\Delta{l}$, which goes from large to small scales.
Specifically, we accumulate the deformations of the grid~$\vec{x}$ by the intermediate results~$\vec{u}_{l_i}=\nabla\times\vec{a}^{\vec{u}}_{l_i}$ over all scales~$l_i$.
We then interpolate the random magnetic vector potential~$\vec{a}^{\vec{b}}(\vec{x})$ at the final deformed grid.
Note that this interpolation is not done intertwined with the scale-by-scale generation of~$\vec{a}^{\vec{b}}(\vec{x})$, but only once at the end of the procedure.
Thus, we are solving the advection equation for the magnetic vector potential
\begin{equation}
    (\partial_t+\vec{u}\cdot\nabla)\,\vec{a}^{\vec{b}}_{l_\mathrm{min}}=0,
    \label{eq:advection}
\end{equation}
which is exact in two dimensions making this approach especially suitable for strong guide field situations~\citep{zank_TheoryTransportNearly_2017}.
Herein lies the key difference with the previous MMLM procedure, which applied the interpolation directly to the field~$\vec{b}(\vec{x})$ in a scale-by-scale fashion.
It should further be noted, that the curl of the intermediate results~$\vec{u}_{l_i}=\nabla\times\vec{a}^{\vec{u}}_{l_i}$ is computed on a uniform grid, so while~$\vec{u}_{l_i}$ serves as an indicator for the grid deformation, it is not affected by it.

The deformation timescale~$\tau=c\,l_i/\operatorname{max}(\|\vec{u}_{l_i}\|)$ is normalized to the maximal value of the current velocity magnitude and governed by the constant~$c$.
This constant is a free parameter, which must be chosen carefully, as it should be large enough for coherent structures to emerge, but not too large to avoid decorrelation.
For higher values of~$c$, energy accumulates on smaller scales, which is corrected by reweighting~$\vec{a}^{\vec{b}}_{l_\mathrm{min}}(\vec{x})$ after interpolation with~$k^{\delta/2}$, where~$\delta$ is the deviation of the energy spectrum scaling from the expected~$-5/3$rd scaling.
Additionally, we mimic the effect of dissipation by applying a low-pass filter~$\exp(-{k^2}/{2k_0^2})$ controlled by the artificial dissipation wavenumber~$k_0$.

For comparison, we perform a direct numerical simulation of incompressible resistive MHD turbulence in a three-dimensional periodic box with resolution~$N^3=1024^3$, no background field and equal diffusivity and resistivity~$\nu=\eta=1.2\times 10^{-3}$~\citep{muller_SpectralEnergyDynamics_2005}.
We obtain Taylor-scale Reynolds numbers $\mathcal{R}_\lambda=439$ for the velocity field and~$\mathcal{R}_{\lambda,m}=94$ for the magnetic field.
The velocity is driven on Fourier modes~$1\le k\le 3$ with the random forcing proposed in~\cite{alvelius_RandomForcingThreedimensional_1999}, which exhibits very low mean cross-helicity and low noise in the time evolution of turbulence bulk quantities.
For this purpose we employed the pseudo-spectral code \textit{SpecDyn}, which was developed in the context of magnetic dynamo action and is tailored for use on modern HPC systems~\citep{wilbert_NumericalInvestigationFlow_2022,wilbert_ImplementationApplicationPseudospectral_2023}.

\section{Results}
We generate sample fields of the continuous cascade process~(CC) given by Equation~(\ref{eq:cascade}) and its minimal multiscale Lagrangian mapping extension~(LM) corresponding to Equation~(\ref{eq:advection}) on a three-dimensional periodic grid with resolution~$N^3=1024^3$.
The parameters of both processes are~$H=1/3$,~$l_0=0.5$ and~$l_\mathrm{min}\approx 1.5\dd{x}$.
The intermittency parameter of the CC process is~$\mu=0.4$, and the additional parameters of the LM process are~$\mu=0.1$,~$c=0.2$,~$\delta=-0.45$, and~{$k_0=256$}.

For visual inspection, slices of field strength and fieldline curvature are plotted in Figure~\ref{fig:spectrum}, together with a plot of the radially averaged energy spectra.
MHD turbulence is characterized by elongated and intricately intertwined coherent structures, while the CC field consists of incoherent intermittent ``clouds'' of large field strength values.
The LM field exhibits thin and intense coherent structures, which are more spatially isolated.
The CC spectrum matches the expected~$-5/3$rd scaling well, the LM spectrum exhibits a bottleneck effect at high wavenumbers, and the MHD spectrum is affected by strong resistivity.
The bottleneck effect of the LM spectrum is due to the simple advective nature of the Lagrangian mapping, which causes accumulation of energy on small scales.
This is counteracted by reweighting the vector potential with~$k^{\delta/2}$ and applying artificial dissipation, however, when balancing the relevant parameters~$c$,~$\delta$ and~$k_0$, a residual bottleneck effect remains.
In order to take potential influences of these different spectra into account, the analysis of curvature and particle transport is also performed with the respective phase-randomized fields.

\begin{figure}
    \centering
    \includegraphics{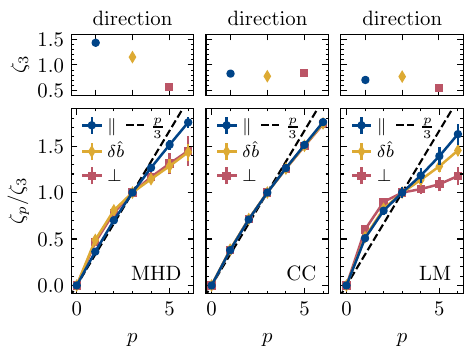}
    \caption{Scaling exponents of conditional structure functions of the magnetic field~$\big\langle\big(\delta b_\perp\big)^p|\,\hat{d}\,\big\rangle\propto d^{\zeta_p}$ for the MHD, CC and LM case.
    The averages of the increments~$\delta{b}_\perp$ are conditional on the displacement vector~$\vec{d}$ being aligned with the local mean field~$\vec{b}_\mathrm{local}$~($\parallel$), the normal fluctuation direction~$\delta\hat{b}_{\perp,N}\perp\hat{b}_\mathrm{local}$~($\delta\hat{b}$), or with the binormal direction~$\delta\hat{b}_{\perp,N}\times\hat{b}_\mathrm{local}$~($\perp$).
    The scaling exponents are normalized as~$\zeta_p/\zeta_3$ and the raw values of~$\zeta_3$ are shown in the top panels.}
    \label{fig:sfscaling}
\end{figure}

We employ conditional structure functions~\citep{mallet_MeasuresThreedimensionalAnisotropy_2016} to study intermittency with respect to the local structure of the field.
See also~\citep{matthaeus_localanisotropy_2012} for a discussion of local anisotropy statistics.
Given a displacement vector~$\vec{d}$ at a point~$\vec{X}$, we consider increments~$\delta\vec{b}_\perp=\vec{b}_\perp(\vec{X}+\vec{d})-\vec{b}_\perp(\vec{X})$ perpendicular to the local mean field~$\vec{b}_\mathrm{local}=\big(\vec{b}(\vec{X}+\vec{d})+\vec{b}(\vec{X})\big)/2$.
Averages over these increments are taken conditionally on the direction of~$\vec{d}$ in an instantaneous local basis given by~$\hat{b}_\mathrm{local}$, the normal direction of the fluctuations~$\delta\hat{b}_{\perp,N}$ with~$\delta\vec{b}_{\perp,N}=\delta\vec{b}_\perp-\big(\delta\vec{b}_\perp\cdot\hat{b}_\mathrm{local}\big)\hat{b}_\mathrm{local}$ and the binormal direction~$\hat{b}_\mathrm{local}\times\delta\hat{b}_{\perp,N}$.
Based on this, the scaling exponents of the conditional averages~$\big\langle\|\delta\vec{b}_\perp\|^p|\,\hat{d}\,\big\rangle\propto d^{\zeta_p}$
are denoted as~$\zeta_{p,\parallel}$,~$\zeta_{p,\delta\hat{b}}$ or~$\zeta_{p,\perp}$
depending on~$\vec{d}$ being aligned with~$\hat{b}_\mathrm{local}$,~$\delta\hat{b}_{\perp,N}$ or~$\hat{b}_\mathrm{local}\times\delta\hat{b}_{\perp,N}$.

Figure~\ref{fig:sfscaling} shows the normalized scaling exponents~$\zeta_p/\zeta_3$ as well as the raw values of~$\zeta_3$ for the three models and the three directions.
The further~$\zeta_p/\zeta_3$ deviates from the linear case, the more intermittent is the process, and smaller values of~$\zeta_3$ correspond to a rougher process.
In the MHD case, the field is the most smooth and non-intermittent parallel to the local mean field direction~$\hat{\vec{b}}_\mathrm{local}$, while it is equally intermittent in the direction of fluctuations~$\delta\hat{\vec{b}}_{\perp,N}$ and the binormal direction~$\hat{\vec{b}}_\mathrm{local}\times\delta\hat{\vec{b}}_{\perp,N}$, and being the most rough in the binormal direction.
In contrast to this, the CC field shows no significant difference between the three directions, which are all equally rough and intermittent, which is expected from the lack of coherent structures.
Lastly, the LM field exhibits again directional dependency, with the parallel direction~$\hat{\vec{b}}_\mathrm{local}$ being the most non-intermittent and the binormal direction~$\hat{\vec{b}}_\mathrm{local}\times\delta\hat{\vec{b}}_{\perp,N}$ being the most intermittent.
However, the field is in the fluctuation direction~$\delta\hat{\vec{b}}_{\perp,N}$ less intermittent compared to the MHD case.

\begin{figure}
    \centering
    \includegraphics{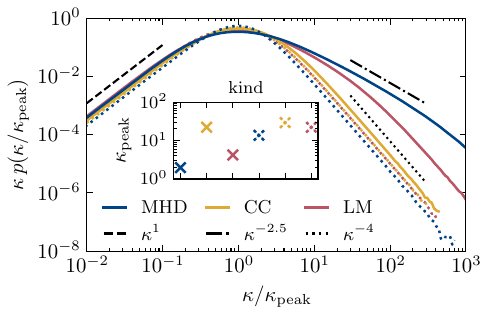}
    \caption{Compensated distributions of fieldline curvature~$\kappa\,p(\kappa/\kappa_\mathrm{peak})$ for the MHD, CC and LM case, as well as their respective random-phase cases (dotted lines).
    The low-curvature~$\kappa^{1}$ scaling, the MHD~$\kappa^{-2.5}$ scaling and the Gaussian~$\kappa^{-4}$ scaling are indicated for comparison.
    The inset shows the respective values of~$\kappa_\mathrm{peak}$}
    \label{fig:curvature}
\end{figure}

While the conditional structure functions provide an extensive statistical picture of magnetic turbulence, additional geometric insight can be gained from the fieldline curvature
\begin{equation}
    \kappa=\|\hat{b}\cdot\nabla\hat{b}\|=\|\hat{b}\times(\vec{b}\cdot\nabla\vec{b})\|/\|\vec{b}\|^2.
    \label{eq:curvature}
\end{equation}
Figure~\ref{fig:curvature} shows the distributions~$p(\kappa)$ for the three cases and their random-phase counterparts.
The MHD case, in agreement with the literature~\citep{yang_RoleMagneticField_2019,yuen_CurvatureMagneticField_2020}, behaves asymptotically as~$p(\kappa)\underset{\kappa\to\infty}{\sim}\kappa^{-2.5}$.
Note, as shown by~\citep{yang_RoleMagneticField_2019}, the scaling becomes~$\kappa^{-2}$ in 2D turbulence, while similar values are expected for general collisionless plasmas.
The CC case agrees, apart from being slightly wider, with the Gaussian random-phase fields, which scale distinctly as~$p(\kappa)\underset{\kappa\to\infty}{\sim}\kappa^{-4}$.
Finally, the LM case appears as an intermediate case between the previous two cases;
its distribution function~$p(\kappa)$ is slightly more narrow than the MHD case and around the slightly right-shifted peak,~$p(\kappa)$ faintly resembles the~$\kappa^{-2.5}$ scaling, before adjusting to the Gaussian~$\kappa^{-4}$ scaling.

The extended flat tail for large~$\kappa$ in the MHD case is caused by a significant amount of intermittent sharp fieldline bends scattered throughout the domain~\citep[see also][]{kempski_CosmicRayTransport_2023,lemoine_ParticleTransportLocalized_2023} and the low value of~$\kappa_\mathrm{peak}$ comes from coherent structures extended on scales comparable to the box size.
In contrast to this, fieldline bends in Gaussian fields are distributed in a self-similar way in the domain, thus leading to a much stepper decay of the distribution.

\begin{figure}
    \centering
    \includegraphics{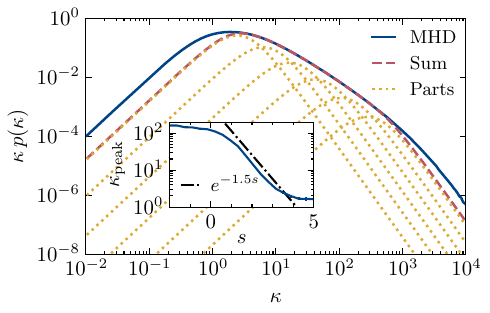}
    \caption{Modelling of the compensated MHD fieldline curvature distribution~$\kappa\,p(\kappa)$ as a weighted sum of shifted Gaussian fieldline curvature distributions.
    The Gaussian fields have power-law spectra~$S(k)\sim k^{-s}$, where the spectral slope~$s$ determines the value of the most dominant curvature~$\kappa_\mathrm{peak}$.
    The weights (not shown) scale analogously to the compensated high-curvature tail with~$\kappa^{-1.5}$.}
    \label{fig:mixture}
\end{figure}

Since the random-phase fields are Gaussian random variables, we expect a universal normalized fieldline curvature distribution~$p(\kappa/\kappa_\mathrm{peak})$, which is independent of the energy spectrum.
This behaviour is observed in numerical experiments and also indicated by~\citep{yang_RoleMagneticField_2019}.
However,~$\kappa_\mathrm{peak}$ does depend on the energy spectrum, e.g.~via the slope~$s$ in case of a power-law spectrum~$\sim k^{-s}$.
Flatter spectra implicate more energy on small scales, resulting in more contributions from high curvatures and consequently larger~$\kappa_\mathrm{peak}$.
This connection is illustrated in Figures~\ref{fig:mixture}, where~$p_\mathrm{MHD}(\kappa)$ is modeled as a weighted sum of Gaussian components, similar to the description of the turbulent velocity increment distribution function as a Gaussian scale mixture~\citep{castaing_1990,lubke_StochasticInterpolationSparsely_2023}.

\begin{figure}
    \centering
    \includegraphics{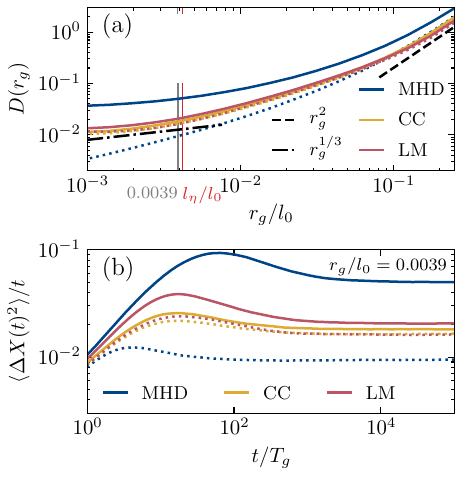}
    \caption{%
    {(a)}~Diffusion coefficients for gyro radii~$r_g/l_0=0.001,\cdots,0.25$ for the MHD, CC and LM case, as well as their respective random-phase cases (dotted lines). High- and low-energy predictions from quasilinear theory are given for comparison.
    The dissipative length scale~$l_\eta$ and the gyro radius of the example in the lower panel are indicated on the abscissa. \\
    {(b)}~Running diffusion coefficients of particles with gyro radius~$r_g/l_0=0.0039$ as an example to illustrate the temporal evolution of the transport.
    }
    \label{fig:rdc}
\end{figure}

In addition to the insight into the structure of the fields gained by the previous steps, we also study the transport of charged particles by numerically solving test particle trajectories~$\vec{X}(t)$ according to the Newton-Lorentz equation
\begin{equation}
    \ddot{\vec{X}}(t)=l_0/r_g\,\dot{\vec{X}}(t)\times\vec{b}\big(\vec{X}(t)\big)
\end{equation}
with a Boris integrator~\citep{ripperda_ComprehensiveComparisonRelativistic_2018} and trilinear interpolation~\citep{schlegel_InterpolationTurbulentMagnetic_2020}.
The magnetic field is normalized to~${\langle b^2\rangle}=1$ and the particles are parameterized by their normalized gyro radius~$r_g/l_0=\gamma mcv_0/qb_0l_0$, where~$l_0$ denotes the outer scale,~$b_0$ denotes the root mean square strength of the magnetic field, and~$v_0$,~$\gamma=1/\sqrt{1-v^2/c^2}$,~$m$ and~$q$ denotes the particle's velocity magnitude, Lorentz factor, mass and charge.

We record mean squared displacements~$\langle\Delta X^2(t)\rangle=\langle\|\vec{X}(t)-\vec{X}(0)\|^2\rangle$ and diffusion coefficients~$D(r_g)=\lim_{t\to\infty}{\langle\Delta X^2(t)\rangle/t}$, once the trajectories have reached diffusive behaviour.
The diffusion coefficients obtained from the three models are plotted in Figure~\ref{fig:rdc}a, and compared with their random-phase counterparts and quasi-linear predictions~\citep{subedi_ChargedParticleDiffusion_2017}.
Figure~\ref{fig:rdc}b shows the exemplary time evolution of~$\langle\Delta X^2(t)\rangle$ at~$r_g/l_0=0.0039$, consisting of an initial super-diffusive phase and short sub-diffusive phase, before arriving at stable diffusive behaviour.
Particles have the largest diffusion coefficients in MHD turbulence on all scales, which stands in striking difference to the random-phase MHD field, where we find the smallest diffusion coefficients.
This behaviour can be explained by the strong deviation of the MHD energy spectrum at high wavenumbers from the~$-5/3$rd spectral slope, and highlights impressively the effectiveness of coherent structures in regard to charged particle transport.
The CC case achieves only a very minor increase compared to the random-phase case, and while the LM case performs better, it is still outperformed by MHD.

\section{Conclusion}
We have presented a novel algorithm for synthetic magnetic turbulence based on a combination of the continuous cascade model, generalized to three-dimensional divergence-free vector fields, and the minimal multiscale Lagrangian map.
The most important differences to previous works on the MMLM procedure are the explicit cascade structure of the underlying noise, primarily working with the vector potential, and interpolating from deformed to uniform grid once at the end of the procedure instead of interpolating at each scale.
These changes considerably strengthen the emergent advective structures.
We compare this algorithm with an incompressible resistive MHD turbulence simulation and the pure three-dimensional continuous cascade model, which is intermittent but lacks coherent structures.
This comparison is done by means of visual inspection, the energy spectrum, conditional structure function scaling, the fieldline curvature distribution and running diffusion coefficients of charged test particles.

We observe that our algorithm produces turbulence exhibiting pronounced coherent structures, albeit not as densely and intricately organized as MHD coherent structures.
This is accompanied by non-trivial conditional structure function scaling, revealing local anisotropy, i.e.~relatively low roughness~($\zeta_{3,\parallel}^\mathrm{LM}>\zeta_{3,\perp}^\mathrm{LM}$) and weak intermittency parallel to the local mean magnetic field, and strong intermittency in the perpendicular direction in agreement with MHD turbulence.
However, when directly compared to the MHD case, the field in the parallel direction is clearly rougher~($\zeta_{3,\parallel}^\mathrm{LM}<\zeta_{3,\parallel}^\mathrm{MHD}$), and the fluctuation direction is not as intermittent as required.
Further, the fieldline curvature distribution resembles the MHD case at small and intermediate curvatures, but exhibits Gaussian scaling at large curvatures.
Finally, while charged particle transport is enhanced, it remains outpaced by the MHD case, as expected due to the simpler geometry and smaller length scales of the synthetic coherent structures.

In conclusion, our algorithm presents significant progress towards simulating realistic turbulence.
Remaining issues are clearly identified and will guide further improvements in designing synthetic turbulence models.
For instance, a feedback mechanism during the algorithm would be highly relevant, which acts on the velocity field~$\vec{u}$ and takes the current state of the deformed grid and the advected magnetic field~$\vec{b}$ into account.
An approach based on the Elsässer formulation of the MHD equations appears promising as well.
Alternatively, one could aim to design a synthetic scalar curvature field, instead of a full vector field, and make use of recent results linking fieldline curvature and charged particle transport~\citep{kempski_CosmicRayTransport_2023,lemoine_ParticleTransportLocalized_2023}.
Such an approach could build on the description of the non-trivial MHD fieldline curvature distribution as a weighted sum of Gaussian components, as presented in this letter.

\acknowledgments{%
J.L. would like to thank T. Schorlepp, P. Reichherzer, A. Schekochihin and P. Lesaffre for helpful discussions.
The authors also thank the two anonymous referees, whose comments helped to improve the manuscript.
This work is supported by the Deutsche Forschungsgemeinschaft (DFG, German Research Foundation) within the Collaborative Research Center SFB1491.
F.E. acknowledges additional support from NASA LWS grant K1327 and DFG grant EF 98/4-1.
The authors gratefully acknowledge the Gauss Centre for Supercomputing e.V. (\url{www.gauss-centre.eu}) for funding this project by providing computing time on the SuperMUC-NG at Leibniz Supercomputing Centre (\url{www.lrz.de}) and through the John von Neumann Institute for Computing (NIC) on the GCS Supercomputers JUWELS at Jülich Supercomputing Centre (JSC).}

\end{document}